\documentclass[prl,twocolumn,showpacs,superscriptaddress,letterpaper]{revtex4-1}

\usepackage{graphicx}
\def \uudd {\uparrow \uparrow \downarrow \downarrow}

\begin{document}
\title{Pressure effects on magnetic ground states in cobalt doped
multiferroic Mn$_{1-x}$Co$_{x}$WO$_4$}

\author{Jinchen Wang}
\affiliation{Department of Physics, Renmin University of China, Beijing 100872, China}
\affiliation{Center for Advanced Materials, Department of Physics and
Astronomy, University of Kentucky, Lexington, Kentucky 40506, USA}
\affiliation{Quantum Condensed Matter Division, Oak Ridge National Laboratory,
Oak Ridge, Tennessee 37831, USA}
\author{Feng Ye}
\email{yef1@ornl.gov}
\affiliation{Quantum Condensed Matter Division, Oak Ridge National Laboratory,
Oak Ridge, Tennessee 37831, USA}
\affiliation{Center for Advanced Materials, Department of Physics and
Astronomy, University of Kentucky, Lexington, Kentucky 40506, USA}
\author{Songxue Chi}
\affiliation{Quantum Condensed Matter Division, Oak Ridge National Laboratory,
Oak Ridge, Tennessee 37831, USA}
\author{Jaime~A.~Fernandez-Baca}
\affiliation{Quantum Condensed Matter Division, Oak Ridge National Laboratory,
Oak Ridge, Tennessee 37831, USA}
\affiliation{Department of Physics and Astronomy, University of Tennessee,
Knoxville, Tennessee 37996, USA}
\author{Huibo~Cao}
\affiliation{Quantum Condensed Matter Division, Oak Ridge National Laboratory,
Oak Ridge, Tennessee 37831, USA}
\author{Wei~Tian}
\affiliation{Quantum Condensed Matter Division, Oak Ridge National Laboratory,
Oak Ridge, Tennessee 37831, USA}
\author{M.~Gooch}
\author{N.~Poudel}
\author{Yaqi~Wang}
\author{Bernd~Lorenz}
\affiliation{
Department of Physics and TCSUH, University of Houston, Houston, Texas 77204,
USA}
\author{C.~W.~Chu}
\affiliation{
Department of Physics and TCSUH, University of Houston, Houston, Texas 77204,
USA}
\affiliation{Lawrence Berkeley National Laboratory, 1 Cyclotron Road,
Berkeley, CA 94720, USA}
\date{\today}

\begin{abstract}
Using ambient pressure x-ray and high pressure neutron diffraction, we studied the pressure
effect on structural and magnetic properties of multiferroic Mn$_{1-x}$Co$_x$WO$_4$ 
single crystals ($x=0, 0.05, 0.135$ and $0.17$), and compared it with the
effects of doping.  Both Co doping and pressure stretch the Mn-Mn chain along
the $c$~direction.  At high doping level ($x=0.135$ and $0.17$), pressure and
Co doping drive the system in a similar way and induce a spin-flop
transition for the $x=0.135$ compound.  In contrast, magnetic ground states at
lower doping level ($x=0$ and $0.05$) are robust against pressure but
experience a pronounced change upon Co substitution.  As Co introduces both
chemical pressure and magnetic anisotropy into the frustrated magnetic system,
our results suggest the magnetic anisotropy is the main driving force for the
Co induced phase transitions at low doping level, and chemical pressure plays
a more significant role at higher Co concentrations.
\end{abstract}


\maketitle

There has been long pursuit for materials showing coupled magnetic and electric
properties, for both technological potential and fundamental scientific
interest. Inspired by the magnetic control of ferroelectric polarization in
TbMnO$_3$ \cite{Kimura03}, considerable interest has focused on the ``type II'' 
multiferroic materials where the ferroelectricity has a magnetic origin 
\cite{Khomskii06,Cheong07,Ramesh07,Tokura07,Tokura14}.  
It can be realized through spin current or inverse Dzyaloshinskii-Moriya 
interaction \cite{Katsura05,Sergienko06,Mostovoy06}, exchange striction \cite{Hur04}, and
$p$-$d$ hybridization mechanism \cite{Jia06,Jia07}.  Magnetic frustration is a
key ingredient in these materials where the competing interactions often lead
to noncollinear spin structure and delicately balanced ground states, and the 
corresponding electric or magnetic properties can be easily tuned by external perturbations.

\begin{figure}[thb!]
    \includegraphics[width=3.2in]{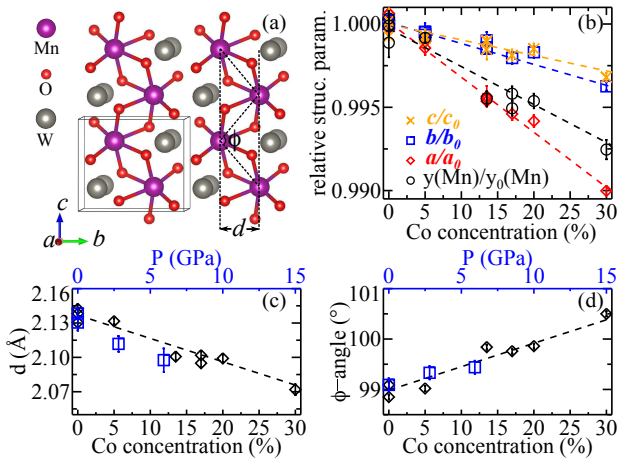}
    \caption{(a) Crystal structure of Mn$_{1-x}$Co$_x$WO$_4$. The unit cell is denoted
	by a box in the lower left corner, and the Mn-Mn chain and its characteristic
	parameters are illustrated. 
	The Co doping dependence of (b) $a$,$b$,$c$ and $y({\rm Mn})$,
	(c) chain displacement $d$ and (d) Mn-Mn-Mn angle $\phi$, are measured at room
	temperature and ambient pressure.
	Data in (b) are normalized to the undoped values to present the relative change.
	In (c) and (d), the pressure effects on the parent compound (blue) were calculated 
	from Ref.~\cite{Macavei93} and compared with Co doping (black).
	}
	\label{str}
\end{figure}

The multiferroic MnWO$_4$ is a classic example of frustrated magnets with
coupled electric and magnetic properties \cite{Taniguchi06,Arkenbout06,Heyer06}.
It crystallizes in the monoclinic wolframite structure (space group $P2/c$).
The edge-sharing MnO$_6$ octahedra form zigzag chains along the $c$ axis
[Fig.~\ref{str} (a)]. When cooling, MnWO$_4$
undergoes successive magnetic transitions \cite{lautenschlager93}.
The incommensurate (ICM) AF3 phase orders below 13.5 K, forming a collinear
sinusoidal structure with a $T$-dependent magnetic wavevector. An ICM
AF2 phase is stabilized between 7 K $<T<$ 12.6 K, with the wavevector locked
at $\vec{q}=(0.214,0.5,-0.457)$, hosting a spiral
magnetic structure that breaks the inversion symmetry \cite{Urcelay-Olabarria13}
and a spontaneous electric polarization along the $b$ axis. Below 7 K,
the electric polarization disappears simultaneously with the AF2 phase. The
commensurate AF1 phase sets in with wavevector $\vec{q}=(0.25,0.5,-0.5)$,
forming a collinear $\uudd$ configuration along the chain.  Both experimental
\cite{Ehrenberg99,Ye11} and theoretical \cite{Tian09,Matityahu12} studies have
revealed sizable long-range magnetic interactions. The system is susceptible
to different perturbations including magnetic field
\cite{Taniguchi06,Taniguchi08,Mitamura12,Urcelay-Olabarria14,Urcelay-Olabarria15}
and chemical doping \cite{Ye08,Meddar09,Chaudhury11,Yu13,Song14,Kumar15}.

Among chemical substitutions with either magnetic or nonmagnetic ions,
the Co-doped system exhibits the most complex magnetic properties
\cite{Song09,Song10,Chaudhury10,Ye12,Urcelay12a,Urcelay12b,Poudel14}.  Only
a few percent of cobalt suppresses the AF1 magnetic structure and stabilizes
the AF2 phase.  Further doping above $x=0.075$ changes the spin structure into
another spiral configuration (AF5 phase) accompanied by polarization flipping
into the $ac$ plane \cite{Ye12,Urcelay12b}. When the Co concentration goes
beyond $x=0.15$, the polarization changes back to the $b$ axis, and the
magnetic structure forms a conical configuration with two modulation vectors
of both AF2 spiral and AF4 collinear components \cite{Urcelay12a,Ye12}. To
understand the successive changes of magnetic and electric polarization
states, Liang {\it et al.}~discussed the coupling between polarization and
magnetic order parameters, which is a high order term in the free energy
expansion \cite{Liang12}. The transition is also discussed as the result of
competing magnetic anisotropy field between Mn and Co ions
\cite{Hollmann10,Song13}. The importance of anisotropy is indeed underscored
by the distinct phase diagrams for different magnetic dopants
\cite{Ye08,Chaudhury11,Song14,Kumar15}. Besides, Co$^{2+}$ also introduces a
different chemical pressure due to its smaller ionic size, which is known as
an important tuning parameter in magnetically frustrated systems.  Earlier
pressure measurements on pure MnWO$_4$ have revealed the evolution of the
crystal structure and a phase transition to triclinic symmetry around 18-25
GPa \cite{Macavei93,Ruiz12,Dai13,Ruiz14,Ruiz15}.  But the pressure effect on
magnetic structures remains unclear.  In this paper, we present the magnetic
phase evolution of Mn$_{1-x}$Co$_x$WO$_4$ under hydrostatic pressure.
Pressure has a pronouced effect at high doping regime, where it induces a
spin-flop transition and drives the system in a strikingly similar manner as
Co doping.  The similarity is also found in the crystal structure evolution,
thus suggesting the chemical pressure takes effect at high Co concentrations.
In the low doping regime however, pressure has limited effect on the magnetic
structures.  The distinct pressure responses reveal interesting contrast
between different Co concentrations.

Single crystals of $\rm Mn_{1-x}Co_xWO_4$ ($0 \le x \le 0.3$) were grown in a
floating zone optical furnace. The chemical compositions were verified by
energy-dispersive x-ray measurements and neutron diffraction refinement
independently. Single crystal x-ray diffraction measurements were conducted at
ambient pressure using a Rigaku XtaLAB PRO diffractometer equipped with a
PILATUS 200K hybrid pixel array detector at the Oak Ridge National Laboratory
(ORNL).  Copper-beryllium cells were used to apply hydrostatic pressures up to
$1.5$ GPa for neutron diffraction experiments. Fluorinert was chosen as the
pressure transmitting medium. The pressure inside the cell was monitored by
measuring the lattice constant of a co-mounted NaCl single crystal.  The
nuclear and magnetic structures were investigated using the four circle
diffractometer HB3A at the High Flux Isotope Reactor (HFIR), ORNL, with
incident wavelength $1.5424~{\AA}$. The samples were also studied using the
triple-axis spectrometer HB1A at HFIR with fixed $E_i=14.6$~meV, and the
single crystal diffuse scattering spectrometer CORELLI at the Spallation
Neutron Source. The sample temperature was regulated using a closed-cycle
refrigerator at HB3A and HB1A, and a liquid Helium cryostat at CORELLI.

We present the room temperature and ambient pressure X-Ray diffraction results
on Mn$_{1-x}$Co$_x$WO$_4$ in Fig.~\ref{str}, and compare with the pressure
effect on pure MnWO$_4$ (Ref.~[\onlinecite{Macavei93}],
[\onlinecite{Ruiz15}]). A systematic change upon doping is evident. The
lattice parameters along all crystallographic directions shrink linearly due
to the smaller Co radius.  Microscopically, the zigzag chain formed by MnO$_6$
octahedron is characterized by the $b$ axis displacement $d$ and the Mn-Mn-Mn
angle $\phi$ [Fig.~\ref{str}(a)].  With increasing the Co concentration, the
Mn ion at Wyckoff position $(0.5,y,0.25)$ moves towards the center as the
value of $y$ decreases. Combined with the decreasing $b$ axis lattice
constant, the chain is effectively stretched along the $c$-axis, signified by
the enlarged $\phi$ angle and reduced displacement in $d$. Such structural
modification due to Co doping, or chemical pressure, could profoundly
influence the exchange interactions between neighboring magnetic ions and lead
to new magnetic ground states as revealed by the neutron diffraction studies
\cite{Ye12}.

Applying hydrostatic pressure has a similar effect on the crystal structure;
the change in the atomic position $y$ of Mn ions upon pressure is
unnoticeable, while the $b$ axis lattice parameter decreases about twice as
much comparing to the other two axes \cite{Macavei93,Ruiz15}. This causes the
interconnected MnO$_6$ chain to distort in the same way as increasing the Co
concentration [Fig.~\ref{str}(c-d)].  To investigate the effect of pressure
induced structural distortion on the magnetic ground state, we performed
systematic, high-pressure neutron diffraction measurements on the doped $\rm
Mn_{1-x}Co_xWO_4$ at $x=$0, 0.05, 0.135, and 0.17. Each has a distinct
spin structure in the $x$-$T$ phase diagram where three spin-flop
transitions occur at $x=0.02$, $x=0.075$, and $x=0.15$.

\begin{figure}[thb!]
\vskip +.3cm
\includegraphics[width=3.1in]{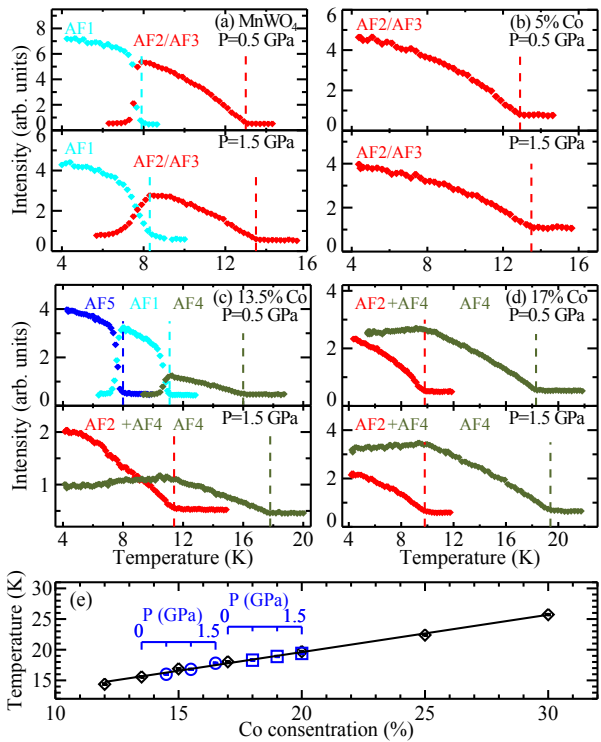}
\vskip -.2cm
\caption{Magnetic order parameters of $\rm Mn_{x}Co_{1-x}WO_4$ at (a)
$x=0$, (b) $x=0.05$, (c) $x=0.135$ and (d) $x=0.17$ at representative pressure
$P=0.5$ and $1.5$~GPa. Measurements of AF5 (blue) and AF2 (red) phases were on
$(-1,0,1)_+$ peaks, AF1 (cyan) phase on $(0,1,0)_+$ peaks and AF4 (green)
phase on $(-0.5,1,1)$ peaks. The onset of transition are marked by vertical
dash lines. (e) Comparison of AF4 transition temperature as a function of
Co-doping (black) and pressure (blue) with a scale of 0.5 GPa/ (Co\%). The
circles are the pressure results from $x=0.135$, and squares are from $x=0.17$. }
\label{ord}
\end{figure}

Figure~\ref{ord} shows the $T$-dependence of the magnetic order parameters
under pressure at four compositions.  At lower doping ($x=0,0.05$),
the sequence of the magnetic transitions remains unchanged while the ordered
moment is gradually reduced with increasing pressure. The transition
temperatures of the AF3 phase for both compositions increase upon pressure at a
rate $\sim 0.45(9)$~K/GPa. The most
prominent change happens at $x=0.135$, where the pressure induces a spin-flop
transition. At $1.5$ GPa, commensurate AF4 peaks persist down to the lowest
temperature, and the collinear AF1 at intermediate temperature and low-$T$ AF5
phase are replaced by the conical spin order normally occurring at higher
Co-doping at ambient pressure \cite{Ye12,Urcelay12a}.  In the case of $x=0.17$, although
the magnetic structure remains the same all the way to $P=1.5$ GPa, the
pressure is gradually enhancing the onset temperature of the AF4 phase and
reducing the ordered moment of the spiral AF2 component, an effect similar as
increasing the cobalt concentration.  All the above mentioned phases at
elevated pressures were quantitative examined by magnetic structure
refinements using collected Bragg peaks intensities \cite{refinementNote}.
For $x=0.135$ and $0.17$, pressure increases the transition temperature
$T_N^{AF4}$ of the collinear AF4 phase at a rate of $\sim 1.2(1)$~K/GPa. 
Doping with Co alone will enhance $T_N^{AF4}$ at a rate of $\sim
0.61(9)$ K for every percent of Co \cite{Ye12,Liang12}. Fig.~\ref{ord}(e)
shows the shift of $T_N^{AF4}$ as function of doping or pressure. Applying
0.5~GPa hydrostatic pressure is equivalent to introducing one percent of Co.
This is consistent with the structural change summarized in
Figs.~\ref{str}(c)-1(d).

\begin{figure}[thb!]
\includegraphics[width=3.3in]{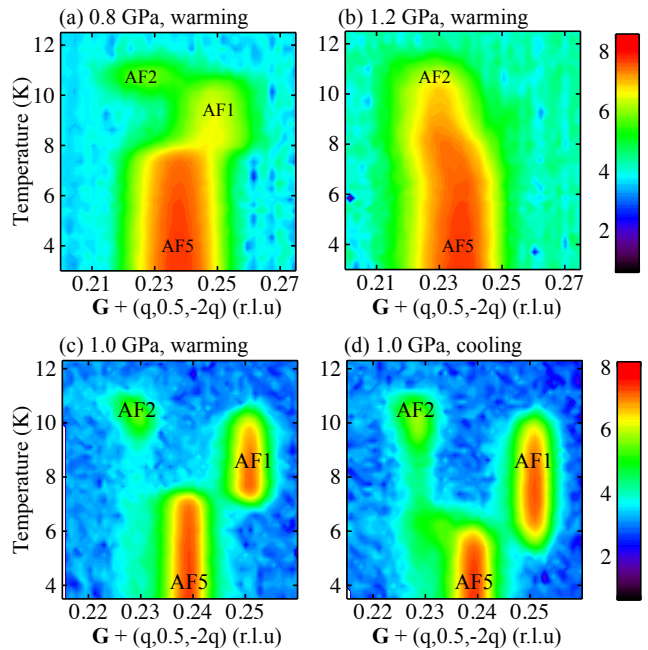}
\caption{The evolution of various magnetic phases of the $x=0.135$ sample
at (a) 0.8~GPa, (b) 1.2~GPa, (c) 1.0~GPa in the warming process and (d) 1.0~GPa in
the cooling process. Panels (a)-(b) and panels (c)-(d) were measured at CORELLI
and HB1A, respectively. For (a) and (b) data are collected near the reciprocal
lattice (rlu) vector $\vec{G}$=(-1,-1,1), which gives magnetic peaks around
$(-0.75,-0.5,0.5)$. Data in (c) and (d) are collected near $\vec{G}$=(0,0,0)
with magnetic peaks near $(0.25,0.5,-0.5)$. }
\label{mesh}
\end{figure}

\begin{figure*}[htb!] 
	\includegraphics[width=5.5in]{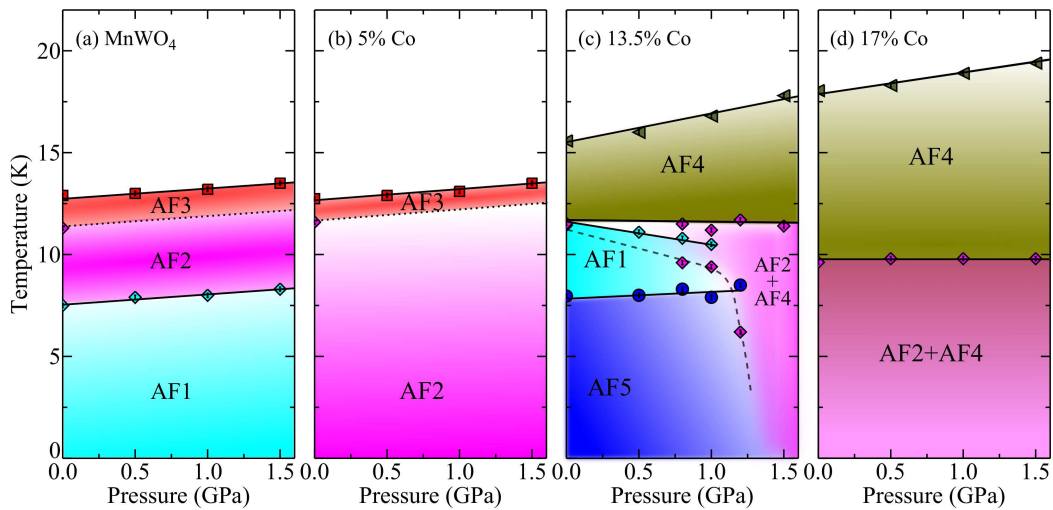} 
	\caption{The $T$-$P$ phase diagrams of $\rm Mn_{1-x}Co_xWO_4$
		determined from neutron diffraction.  Cyan, magenta, orange,
		brown, and blue represent AF1-AF5 phases, respectively.  The dash line in
		panel (c) denotes the lower boundary of AF2 phase.}
	\label{pd} \end{figure*}

The $x=0.135$ sample has the most complex pressure-induced magnetic
transitions among the four compositions. Figure~\ref{mesh} shows the detailed
measurement revealing how the magnetic configuration evolves with increasing pressure.
There is a small trace of AF2 phase around 11 K at ambient pressure
(Fig.~10 of Ref.~[\onlinecite{Ye12}]). It appears in a narrow temperature window of 0.5~K and
gives way to AF1 order when the system is further cooled.  The AF5 order with
an $ac$-spiral becomes the ground state at the lowest temperature.  As the
pressure increases, AF2 order expands in a wider temperature range as well as
the AF5 phase. This shrinks the temperature window of the collinear AF1 phase
as the upper(lower) boundary decreases(increases). As the pressure further
increases to $1.2$ GPa, the AF1 phase is completely suppressed and the AF2
order extends down to lower $T$ and becomes the remaining state that competes
with the AF5 phase.  A strong hysteresis between cooling and warming is
observed at 1.0~GPa [Figs.~\ref{mesh}(c)-3(d)]. The AF2 phase tends to compete
for the ground state during cooling process.  The hysteresis suggests that 
the AF2 to AF5 transition is first order in nature,
therefore clearly signifies the competition of free energy in different
phases.  The pressure lowers AF2's energy and eventually induces the magnetic
transition at a critical point. The corresponding bulk polarization
measurements \cite{Gooch16} on the same batch of $x=0.135$ sample show
excellent agreement with the diffraction results; at low pressure, $P_b$
(associated with the AF2 order) initially shows an expansion of temperature
window near 11~K and is then suppressed with occurrence of $a$ axis
polarization (from AF5) at low $T$.  With increasing pressure, the $b$ axis
polarization gradually takes over and finally at 1.5~GPa reaches the size of
the $x=0.17$ sample at ambient pressure.

Figure~\ref{pd} illustrates the $T$-$P$ phase diagram at different Co
compositions.  Pressure has more significant effects in the high doping
samples ($x=0.135,0.17$) than the low ones ($x=0,0.05$).  At $x=0.135$, the
pressure-induced spin-flop transition is observed.  One might speculate if
the transition is due to the fragility of AF5/AF1 phases or the
proximity of the $x=0.135$ sample to the multi-phases boundary.  The AF1 phase
in the undoped MnWO$_4$ is apparently stable against pressure up to 1.5~GPa
[Fig.~\ref{pd} (a)], while doping a small mount of Co can quickly suppress it.
The $x=0.135$ sample is indeed near the phase boundary where the AF5 changes
to the conical phase.  But it is similar for the $x=0.05$ sample near the
boundary that separates the AF2 and AF5 states, and the undoped sample that is
close to the $x=0.02$ boundary. If the same scale of 3\% additional Co at 1.5
GPa is adopted, we expect the AF1 phase at $x=0$ should be completely
suppressed and the $x=0.05$ sample is driven to the AF5 phase.  However, such
a pressure induced transition does not occur in these samples.  It underscores
distinct pressure responses between the low and high Co concentrations.

The distinct pressure responses of Mn$_{1-x}$Co$_{x}$WO$_4$ reveal considerable
differences in the magnetic Hamiltonian at the low and high Co
concentration regime. The single-ion anisotropy of the Co$^{2+}$ (in
$3d^7$ state) measured from the X-Ray absorption spectroscopy is much
pronounced than the Mn$^{2+}$ (in $3d^5$ state) with half filled $3d$ orbitals
\cite{Hollmann10}, and is likely to provide the largest energy change in the
Hamiltonian. When Co is initially introduced, the overwhelming change in the
single-ion anisotropy induces a drastic change in the magnetic ground state.
This is manifested by the quick suppression of AF1 commensurate state and the
gradual rotation of AF2 spiral plane.  The long-range isotropic exchange
coupling term with the energy scale of 2-3 meV is highly frustrated and
involves intra- and inter-chain interactions ranging from $J_1$ to $J_{11}$
\cite{Ye11}.  The Dzyaloshinskii-Moriya interaction helps to stabilize the
ground state in the parent compound and is important for the inversion
symmetry breaking, but its overall energy scale is orders of magnitude weaker
compared to other terms \cite{Solovyev13}.
The competition between these interactions critically depends on the
structure details such as bond distances and the Mn-O-Mn bond angle.  
It is desirable to investigate the detail changes of Hamiltonian
upon doping and pressure in the future inelastic neutron scattering work.

The consistent scale factor of the structural change [Fig.~\ref{str}(c)-(d)],
the AF4 ordering temperature [Fig.~\ref{ord}(e)], and the pressure induced
transitions at $x=0.135$ suggest a close correlation between the doping and
pressure at high Co concentrations.  Once the Co anisotropy dominates and the
moment direction is locked along the Co easy axis, the magnetic structure is
mainly governed by the perturbation of exchange couplings. The doping- or
pressure-induced distortion in the form of chain
stretching takes over and becomes the main driving force.  This occurs at $x > 0.12$ 
where the AF4 phase sets in. Further doping does not change the 
spin easy direction although Co and Mn still compete in anisotropy
\cite{Herrero15}, and pressure modifies the magnetic states in a
similar way as doping.

In summary, we have performed a systematic pressure study characterizing the
magnetic order of multiferroic Mn$_{1-x}$Co$_{x}$WO$_4$ using x-ray and neutron
diffraction. We found the pressure has a significant effect on spin structure at
high cobalt concentrations and induces a spin-flop transition at $x=0.135$.  In
contrast, the magnetic states at lower doping are rather stable against
pressure.  Our results reveal the balance between controlling the magnetic
anisotropy and modifying the long-range magnetic interactions through
structural change determines the ground state spin order, and offers a viable
method to design magnetoelectric materials with desired properties.

Research at ORNL's HFIR and SNS was sponsored by the Scientific User
Facilities Division, Office of Basic Energy Sciences, U.S. Department of
Energy.  Work at Houston is supported in part by the T.L.L. Temple Foundation,
the John J. and Rebecca Moores Endowment, and the State of Texas through
TCSUH, the US Air Force Office of Scientific Research, Award No. FA9550-
09-1-0656. J.C. Wang acknowledges support from China Scholarship Council.

%

\end{document}